\documentclass{jfm}
\usepackage{graphicx}
\usepackage{epstopdf, epsfig}

\usepackage{subcaption}

\usepackage{float} 

\usepackage{xcolor}
\usepackage[font=small]{caption}
\usepackage[labelformat=empty, position=top]{subcaption}
\usepackage[export]{adjustbox}
\usepackage{anyfontsize}
\usepackage{amsmath}
 \numberwithin{equation}{section}

\shorttitle{Study of bifurcation in coalscence of bubbles using front-tracking method}
\shortauthor{Mehrdad Izadjoo}

\title{Study of bifurcation in coalscence of bubbles in turbulent flows}

\author{Seyedmehdi Abtahi, Mehrdad Izadjoo\aff{2}, 
  \corresp{\email{merdad.izadjoo@gmail.com}}}

\affiliation{\aff{1} University of Illinois at Chicago,  \aff{2}Darab University }

\begin{document}

\maketitle

\begin{abstract}
There are many applications of multiphase flow in important fields such as biological, chemical and power processes. Bubble coalescence is of a significant importance in simulating multiphase fluid flows. Weber number ($W_e$), Reynolds number (Re) and collision parameter play important role in the coalescence of bubbles. In the present work, front-tracking method is applied to simulate bubble coalescence. Moreover, the results are presented for different collision parameters and changes in the coalescence of bubbles are discussed.

\end{abstract}


\section{Introduction}\label{sec:rules_submission}
Bubbly flows play a prominent role in physical, chemical and biological processes, all of which invlove bubble-bubble and bubble-bubble interactions. Bubble coalescence is of a significant importance in determining the interfacial area, thereby affecting the mass and heat transfer between the two phases. Knowledge of the coalescence of two bubbles is consequential to better comprehend the bubbly flow behavior which results in more efficient designs in multiphase flows \citep{esfahanian2014fluid}. 
During coalescence, two drops combines through the liquid bridge between them, then it grows up to the size of drops. Many researchers have studied the dynamics of coalescence of bubbles \citep{chi1989theoretical, basaran1992}.  The coalescence of two bubbles rising in a vertical line is studied by \citep{ramirez2011, esfahanian2014simulation}. The colescence consists of the following steps: the collision of bubbles, thinning of a liquid film, and rupture of the film at some critical thickness \citep{chen2013joint}. To capture trailing bubbles and coalescence behavior, approaching and coalescence of bubbles is significant \citep{hasan2011review}. Liquid viscosity can influence the wake effect of the leading bubbles which results in bubbles interactions \citep{abtahi2020machine1}. Researches on two-rising bubbles has been performed for high and low Reynolds number \citep{rushton1978slow, yuan1994line}. The steps of coalescence after collision is difficult to study 
because the rapture of liquid is so complex that mechanisms are derived by simplifying the coalescence model. Interaction and coalescence between a bubble and a free surface have been studied numerically and experimentally \citep{dehrouyeh2016parametric}. Coalescence of deformable bubble is also carried out \citep{abtahi2020integrated}.

Studies have shown that the coalescence time is not a constant value. By using a binary coalescence cell, the stability of foam films was analysed \citep{dehrouyeh2019nonlinear}. Research showed that a surfactant at an interface implement repulsive forces and stabilizes bubbles against coalescence \citep{kumar2006adsorption}.

Over the past decades, computational fluid dynamics (CFD) has been employed as a powerful tool to simulate multiphase and turbulent flows \citep{peyvan2016axial,abtahi2020turbulent,li2019compressibility}. Many methods such as VOF model \citep{abtahi2020machine}, the level set method \citep{lakdawala2014dual} and lattice Boltzmann method (LBM) \citep{cheng2010retracted} have been applied to simulate bubbles interactions. The VOF model has widely applied to simulate formation, break up and coalescence of bubbles. It is dependent on mesh size, although it is accurate and simple\citep{li2019compressibility}. 

In this study, front-tracking method is employed to simulate the coalescence of bubbles. Moreover, results will be presented for different Weber numbers ($ We $) and collision parameters. It is shown that by changing Weber number and collision parameter, the coalescence behavior will change. The rest of this paper is structured as follows: front-tracking method is explained in section 2. Results of simulations are shown in section 3 and they will be discussed. Conclusion is provided in section 4.

\begin{figure}
  \centerline{\includegraphics{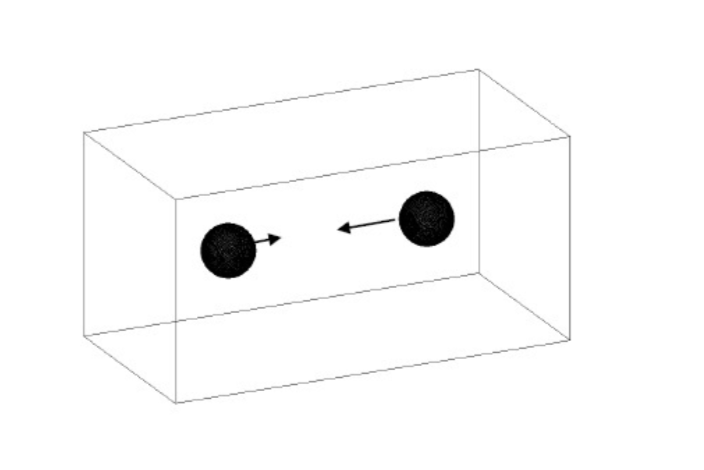}}
  \caption{Geometry of two bubbles which collide to each other.}
\label{fig:ka}
\end{figure}

\begin{figure}
  \centerline{\includegraphics[width=150mm]{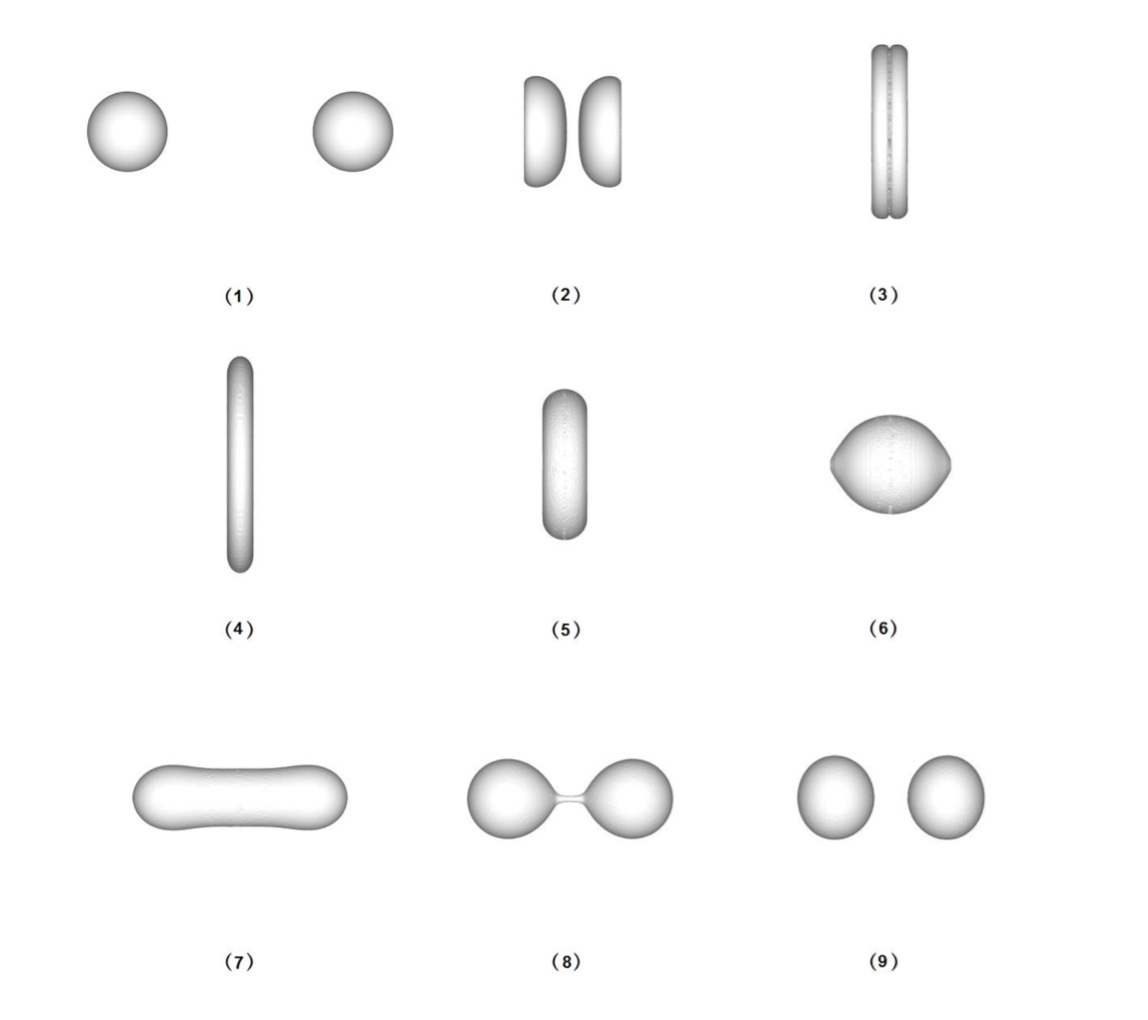}}
  \caption{Reflexive separation of bubbles with I = 0 in t = 0,t = 2.61,t = 3.86,t = 5.42,t = 9.29, t = 11.07, t = 13.88, t = 21.5, t = 25.37.}
\label{fig:kd}
\end{figure}

\begin{figure}
  \centerline{\includegraphics[width=150mm]{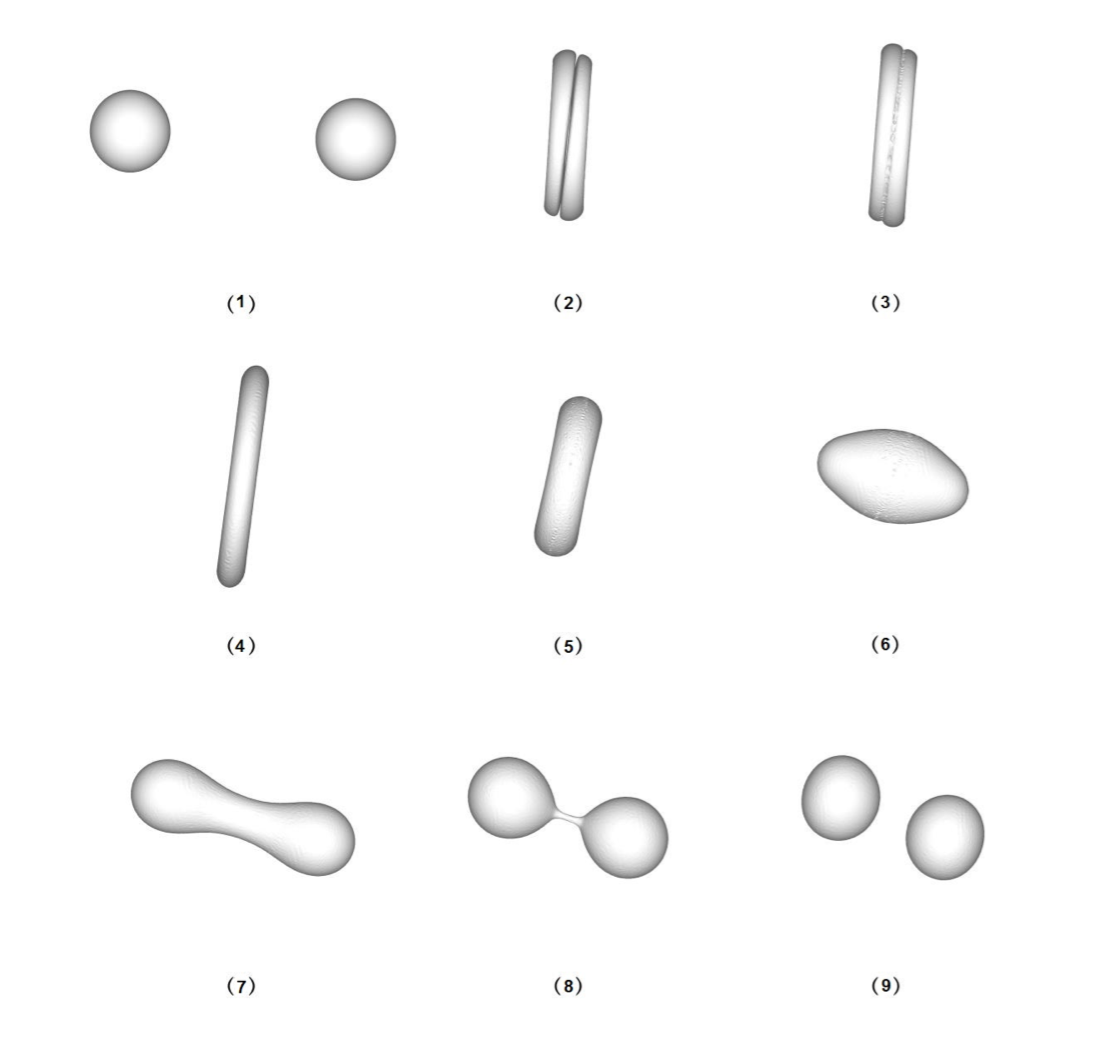}}
  \caption{Reflexive separation of bubbles with I = 0.1 in t = 0,t = 3.65,t = 3.86,t = 5.53,t = 9.08, t = 11.69, t = 16.49, t = 21.5, t = 25.57.}
\label{fig:kd}
\end{figure}

\begin{figure}
  \centerline{\includegraphics[width=150mm]{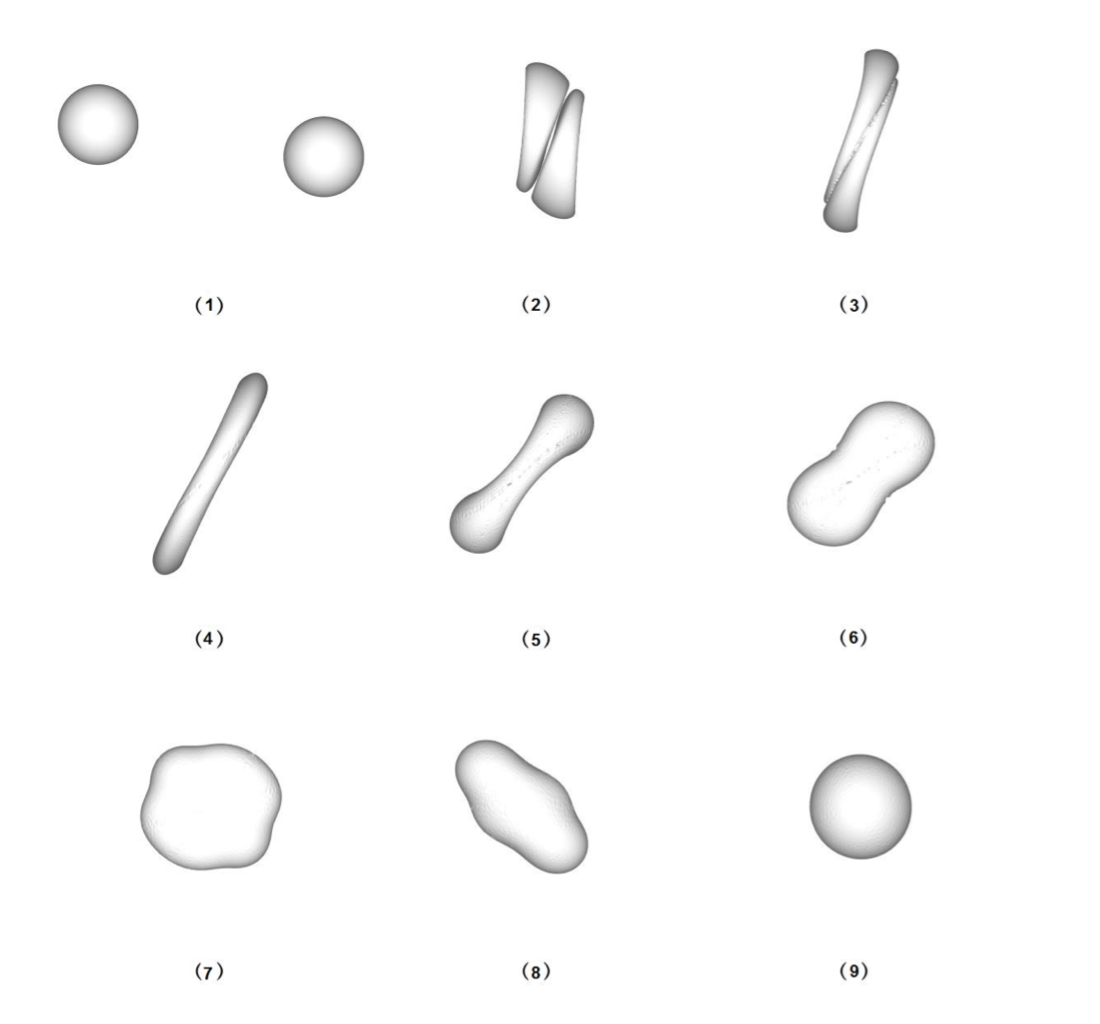}}
  \caption{Permanent coalescence of bubbles with I = 0.4 in t = 0,t = 3.23,t = 4.07,t = 5.64,t = 8.98,t = 10.34,t = 11.8,t = 13.26,t = 38.41.}
\label{fig:kd}
\end{figure}

\begin{figure}
  \centerline{\includegraphics[width=150mm]{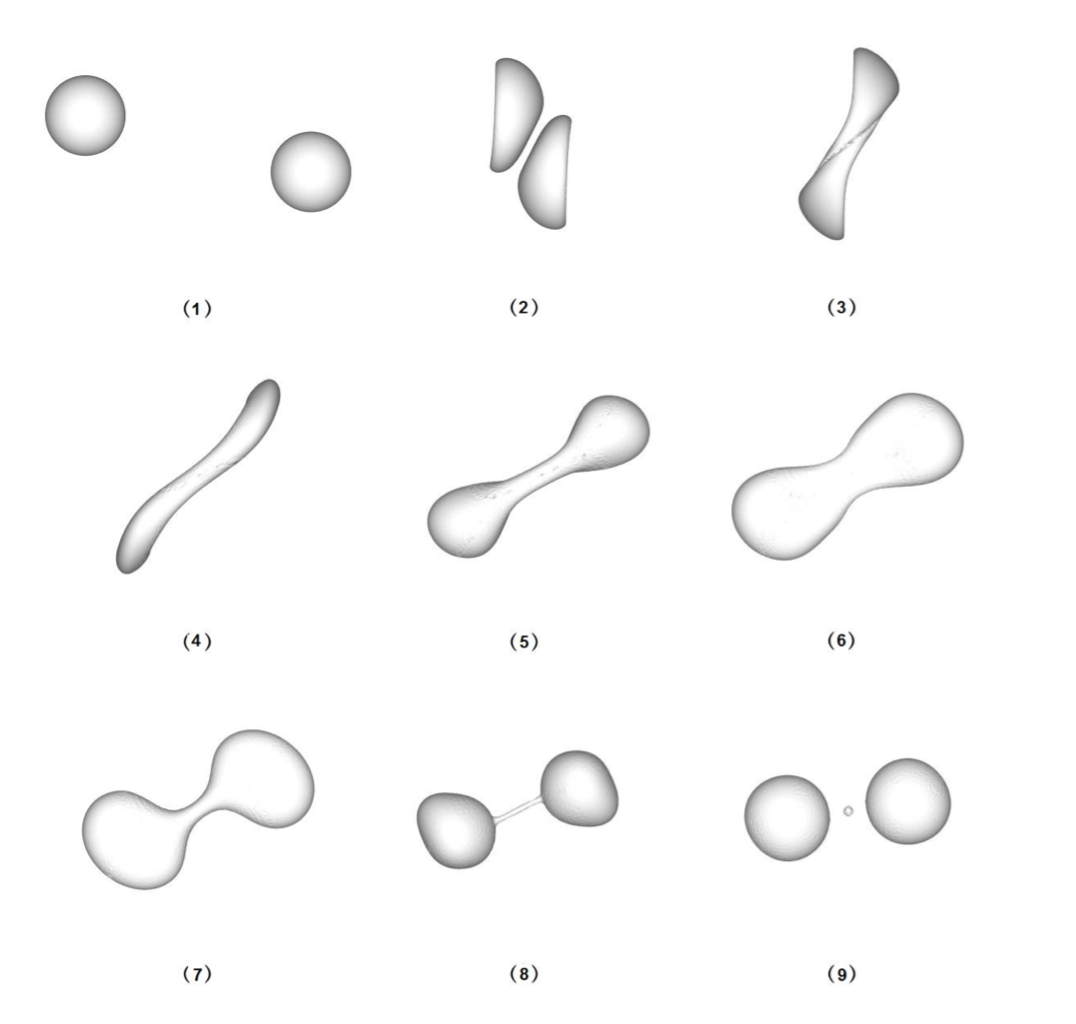}}
  \caption{Stretching separation of bubbles with I = 0.7 in t = 0,t = 2.92,t = 4.07,t = 6.05,t = 8.56,t = 9.5,t = 10.96,t = 13.26,t = 18.37.}
\label{fig:kd}
\end{figure}

\section{Front-tracking method}\label{sec:types_paper}
The Navier-Stokes equations are numerically solved for both gas and liquid phases:
\[\frac{\partial \rho}{\partial t} + \nabla.(\rho VV) = -\nabla P + \rho g + \nabla.\mu_{eff}(\nabla V + V \rangle^T) + \oint_\Delta S \sigma_{eff}kn(r-r_f) \,da \]  
where $V$, $\rho$ and $P$ are normalized velocity, density and pressure, respectfully.
The variables are normalized with characteristic bubble velocity, $V_C$, liquid density $\rho$, the dynamic pressure $\rho_c a V^{2}_c$ and t is normalized with $L_c/V_c$, where $L_c$ is the characteristic length. $\mu_{eff}$ is the reciprocal of Reynolds number, $Re = \frac{\rho V_c L_c}{\mu_l}$, where $\mu_l$ is the liquid viscosity. $\sigma_{eff}$ is the reciprocal of Weber number, $W_e = \frac {\rho V^{2}_c L_c}{\sigma}$, where $\sigma$ is the surface tension of liquid. k is twice the mean curvature, n is the outwardly directed unit normal vector at the bubble surface, and r is the space vector with the subscript f designating the interface. By considering both fluids to be incompressible, continuity can be written as below:
\[\nabla . V = 0\]
Moreover, density and viscosity stay constant for incompressible fluid, $\frac{\partial \rho}{\partial t} = 0$ and $\frac{\partial \mu}{\partial t} = 0$.
 
\section{Result}\label{sec:filetypes}
In the present work, we simulate coalescence of two bubbles in three dimension, which are shown in Figure 3. Regimes of coalescence of bubbles are mainly dependent on $W_e$ and collision parameter, which is defined as $I = \frac{X}{ds}$, where X is the vertical distance between the velocity of center of mass of bubbles and d is the diameter of bubbles. The first regime happens when the two bubbles collide, but they do not have sufficient energy to become separate again. The other important one is reflexive separation, in which the collided bubble separates in three steps. At first, the collided bubble grows in vertical direction and shape as a disk. Then, because of surface tension and curvature difference in longitudinal and radial directions, the bubble changes back to be a sphere. The inertia of bubble makes an longitudinal expansion. Another regime is stretching separation, in which the two bubbles rub each other and after being combined, they separate from each other. Reflexive and stretching separation happen in low and high collision parameters, respectfully.
In the simulation, we consider $W_e$ = 33 and Re = 163.1. Size of solution domain is $1 \times 1 \times 2$, while there are 96 cells in each unit. Moreover, the height of channel is 8 times larger than the radius of bubble. The result for reflexive separation and I = 0 is shown for different time steps in Figure 3.
The result for reflexive separation and I = 0.1 is shown for different time steps in Figure (3). As it is shown in Figure 3, the bubble grows in a different angle and separate in a different angle than the case I = 0.
Permanent coalescence is shown in Figure 3 with I = 0.4, which indicates that the bubble do have the sufficient energy to increase length in vertical direction.
Figure 3 indicates the results for stretching separation with I = 0.7. It must be noticed that in this case a new bubble appears between the two separated bubbles. As it is shown in Figures 3 to 3, by changing I, the coalescence of bubbles will differs (bifurcation in the system).
\section{Conclusion}
In the present study, front-tracking method is applied to simulate the coalescence of bubbles. Collision parameter play an important role in coalescence. Different collison parameters result in reflexive separation, permanent coalescence and stretching separation. In each case results was shown for different time steps to show how the bubble change the shape during coalescence to finally become either separated into two or three bubbles or stay united as a single bubble. By changing the collision parameter, the results differs, which indicates bifurcation in the system.

\bibliographystyle{jfm}
\bibliography{jfm-instructions}

\end{document}